\documentclass[11pt,a4paper]{article}
\pdfoutput=1
\usepackage{graphicx}
\usepackage{amssymb}
\usepackage{amsmath}
\usepackage{url}
\usepackage{graphicx}
\usepackage{amsmath}
\usepackage{amssymb}
\usepackage{color}

{\end{list}}
\newcounter{enumct}

\newlength{\abstwidth}
\begin{document}
\sloppy
 
\pagestyle{empty}
 
\begin{flushright}
LU TP 15-35\\
MCnet-15-26\\
September 2015
\end{flushright}

\vspace{\fill}

\begin{center}
{\Huge\bf Hard Diffraction in 
\textsc{Pythia}~8\let\thefootnote\relax\footnotetext{Presented at the 16th conference on Elastic and Diffractive scattering (EDS Blois 2015)}}\\[10mm]

{\Large Christine O. Rasmussen} \\[3mm]
{\it Theoretical Particle Physics,}\\[1mm]
{\it Department of Astronomy and Theoretical Physics,}\\[1mm]
{\it Lund University,}\\[1mm]
{\it S\"olvegatan 14A,}\\[1mm]
{\it SE-223 62 Lund, Sweden}
\end{center}

\vspace{\fill}

\begin{center}
\begin{minipage}{\abstwidth}
{\bf Abstract}\\[2mm]
We present an overview of the options for diffraction implemented in 
the general--purpose event generator \textsc{Pythia}~8. We review the 
existing model for low-- and high--mass soft diffraction and present 
a new model for hard diffraction in $\mathrm{pp}$ and 
$\mathrm{p}\bar{\mathrm{p}}$ collisions. 
Both models uses the Pomeron approach pioneered by Ingelman and 
Schlein, factorising the single diffractive cross section into a 
Pomeron flux and a Pomeron PDF. The model for hard diffraction is 
implemented as a part of the multiparton interactions framework, 
thereby introducing a dynamical rapidity gap survival probability 
that explicitly breaks factorisation.
\end{minipage}
\end{center}

\vspace{\fill}

\phantom{dummy}

\clearpage

\pagestyle{plain}
\setcounter{page}{1}

\section{Introduction}
While most phenomena in high--energy hadronic collisions have been 
explained by QCD, the effects of the softer hadronic collisions 
remains a mystery. We observe these collisions in experiments, and 
can motivate why they should be present, but the explanation of how 
they occur is still largely based on phenomenological models. These 
models should be able to describe all aspects of such collisions, 
like differential cross sections, one--particle distributions and 
global event characteristics. The models should also describe the 
exclusive topologies of these softer collisions, specifically the 
occurrence of rapidity gaps.

Many models, including the models used in \textsc{Pythia}~8
\cite{Sjostrand:2014zea}, are 
based on Regge theory. In this theory, poles in the plane of 
complex spin $\alpha$ can be seen as hadronic resonances. These 
appear to lie on linear trajectories, $\alpha(t) = \alpha(0) + 
\alpha't$. Most important for high-energy collisions is the 
Pomeron ($\mathbb{P}$) trajectory, with its $\alpha(0) > 1$ 
explaining the rise of the total cross section. 
This state is a colour-singlet carrying the quantum numbers of 
the vacuum. From a modern viewpoint it (predominantly) consists 
of gluons and could thus be called a glueball, or a gluonic 
ladder if present in the final state (a cut Pomeron). A 
topological expansion can be defined, with increasingly complex 
processes. The simplest possible exchange is a 
single--Pomeron one, which gives rise to elastic scattering. 
Multi--Pomeron exchange is also possible, e.g.\ involving the 
triple--Pomeron vertex. This way various diffractive topologies 
can be constructed. In this paper we focus on the single diffractive 
(SD) topologies, since these have the largest diffractive cross 
section and form the starting point on the road towards more 
complex configurations.

Ingelman and Schlein \cite{Ingelman:1984ns} 
proposed a model in which the exchanged 
Pomeron can be viewed as a hadronic state. This opened up the 
possibility for using Pomeron parton distribution functions (PDFs) 
to be combined with a probability for taking out a Pomeron from 
the initial hadronic state, the Pomeron flux. The diffractive system 
can be viewed as a hadron--hadron collision at reduced energy, and 
existing hadron--hadron event generators can be used for modelling 
the diffractive events. The simplest model does not allow for 
multiparton interactions (MPIs), however, or equivantly for the 
final--state effects of multiple cut Pomerons. These MPIs create 
additional colour strings in the event, each string giving rise to 
hadronic production. Hence we risk filling up the rapidity gap 
created by the exchange of the `first' Pomeron. As a rapidity gap 
is needed to trigger on diffractive events, we risk losing a large 
fraction of the could--have--been diffractive events by these MPIs. 
This introduced the concept of rapidity gap survival probability 
(RGSP), which is unique to hadron--hadron collisions, given credibility 
by the lower observed rate of hard diffractive processes at the 
Tevatron than expected from HERA flux/PDF determinations 
\cite{Affolder:2000vb}.

\section{Soft diffraction in \textsc{Pythia}~8}
The soft diffraction machinery available in \textsc{Pythia}~8 was 
originally developed for \textsc{Pythia}~6 \cite{Sjostrand:2006za}, 
but rewritten and expanded 
for the new version, and now includes both single--, double-- and 
central--diffractive systems (SD, DD, CD) as well as elastic 
collisions and non--diffractive topologies \cite{Navin:2010kk}. 
The total hadronic cross 
section is calculated using the Donnachie--Landshoff parametrisation 
\cite{Donnachie:1992ny}, with a Pomeron and Reggeon term. 
The elastic and diffractive 
cross sections are based on the Schuler--Sj\"o{}strand model 
\cite{Schuler:1993wr} 
and the non--diffractive cross section is inferred from these two 
models.

The Schuler-Sj\"o{}strand model is also based on Regge theory and 
gives an approximate $dM/M^2$ mass dependence as well as an 
exponential $t$ dependence. Fudge factors have been introduced 
to the model, to dampen the cross sections close to the kinematical 
limits, as well and to dampen the DD cross section where two diffractive 
systems overlap. Other Pomeron--flux models have also been implemented in 
\textsc{Pythia}~8 (see the manual \cite{PythiaWeb}). 
The subsequent 
hadronisation of a diffractive system is a separate chapter, and the 
same for all Pomeron--flux models. Particle production depends 
strongly on the mass of the diffractive system, however, and hence 
it has been split into two regions.

\subsection{Low--mass soft diffraction}
In the low--mass regime, $M \leq 10$ GeV, energies are not sufficiently 
high to apply a perturbative framework to the Pomeron--proton 
subcollision. Instead we visualise the event as an interaction where 
the Pomeron has ``kicked out'' a parton from the diffractively excited 
hadron. If a valence quark is kicked out then a single colour string 
is stretched between it and the diquark remnant. A kicked--out 
gluon gives a hairpin string topology, stretching from one quark 
in the proton remnant to the gluon and then back to the remaining 
diquark of the remnant. The probability for the Pomeron to interact 
with either a quark or a gluon is mass--dependent, 
$P(q)/P(g) = N/M^p$ with $p$ being a tunable parameter, making 
the gluons dominate at higher mass. There are no additional 
MPIs in the low--mass regime. The strings are hadronised using the 
Lund string fragmentation model \cite{Andersson:1983ia} and gives rise to 
low--$p_T$ activity in the diffractive system.

\subsection{High--mass soft diffraction}
In the high--mass regime, $M > 10$ GeV, a perturbative description 
is attempted. So as not to give any discontinuies, and possibly 
also representing a real physics evolution, the fraction of 
perturbative events gradually increases with $M$ and dominates for 
$M > 20$ GeV. 

In the new component the Pomeron is viewed as a particle 
with partonic content a la Ingelman and Schlein. Thus, once $M$ and 
$t$ have been selected, the system is set up as a 
$\mathbb{P}\mathrm{p}$ collision and a semi--hard perturbative 
$2\rightarrow2$ partonic interaction is selected by the MPI machinery. 
Inside the Pomeron--hadron system the full interleaved evolution of 
initial-- and final--state showers (ISR and FSR) and MPI is applied 
using the Pomeron PDFs. The MPI activity 
in the subsystem has been tuned to give approximately the same amount 
of activity as in non--diffractive events of the same mass, by 
introducing an effective total Pomeron--proton cross section. This 
(tunable) total cross section is set to a constant value of 10 mb, 
slightly higher than other numbers found in the literature. The colour 
strings obtained in the evolved diffractive system are hadronised using 
the Lund string fragmentation model. Jets can be produced in the 
$2\rightarrow2$ partonic processes.

Although the models for soft diffraction available in 
\textsc{Pythia}~8 are largely successful, some minor issues show up. 
Not all aspects of the data are described using the default 
model and settings, both on the level of differential cross sections 
and on that of particle spectra. A retune of parameters used in 
the default model could fix some issues, in particular if allowing 
for more flexible shapes e.g.\ for the Pomeron flux. We intend to 
improve the default models in the near future. 

\section{The new model for hard diffraction in \textsc{Pythia}~8}
The model described above does allow for QCD $2\rightarrow2$ processes 
at all $p_T$ scales, but is primarily intended for lower $p_T$ values. 
It is not intended for the study of truly hard processes, either in 
QCD or beyond. Instead a model for hard diffraction has been developed 
\cite{Rasmussen:2015} based on the assumption 
that the proton PDF can be 
separated into a non--diffractive and a diffractive part, with the 
diffractive part described using the factorisation approach,
\begin{align}
f_{i/\mathrm{p}}(x, Q^{2}) &= f_{i/\mathrm{p}}^{\mathrm{ND}}(x, Q^{2}) + 
  f_{i/\mathrm{p}}^{\mathrm{D}}(x, Q^{2}),\nonumber\\
f_{i/\mathrm{p}}^{\mathrm{D}}(x, Q^{2}) &= \int_{x}^{1} \int_{t_{min}}^{t_{max}} 
  \mathrm{d}t~\mathrm{d}x'~\mathrm{d}x_{\mathbb{P}}~
  f_{\mathbb{P}/\mathrm{p}}(x_{\mathbb{P}}, t)~
  f_{i/\mathbb{P}}(x', Q^{2})~\delta(x - x'x_{\mathbb{P}})\nonumber\\
&= \int_{x}^{1} \frac{\mathrm{d}x_{\mathbb{P}}}{x_{\mathbb{P}}}~
  f_{\mathbb{P}/\mathrm{p}}(x_{\mathbb{P}})~ 
  f_{i/\mathbb{P}}(\frac{x}{x_{\mathbb{P}}}, Q^{2}).
\end{align}

The probability of diffraction on one side is then given as the 
ratio of diffractive to inclusive PDFs,
\begin{equation}\label{Prob}
P^{\mathrm{D}}(x_i,Q^2) = f_{i/\mathrm{p}}^{\mathrm{D}}(x_i,Q^2)
  /f_{i/\mathrm{p}}(x_i, Q^2).
\end{equation}
At high energies most interactions occur at low $x$ where 
$P^{\mathrm{D}}(x, Q^2)\sim 0.1$. Hence we expect approximately 
10--15\% of the events to be diffractive based only on Eq. \ref{Prob}.

In addition the model implements a dynamical gap survival. This means 
we do not allow any further MPIs to occur between the two 
incoming hadrons, so as to ensure the gap survives. 
In practise the tentative classification as diffractive, based on 
Eq. \ref{Prob}, initially has no consequences: all events are handled 
as non--diffractive hadron--hadron collisions. Only if no additional 
MPIs occur does a diffractive classification survive and only then 
is the $\mathbb{P}\mathrm{p}$ 
subsystem set up. A full evolution of ISR, FSR and MPIs is 
performed in this $\mathbb{P}\mathrm{p}$ system, along with 
hadronisation of the colour strings 
in the event. At this stage all non--diffractive events can be 
discarded for a pure diffractive sample, or can be 
kept and hadronised as usual for an inclusive sample. 

The restriction on the number of MPIs in the hadron--hadron system 
introduces an additional 
suppression factor of $\sim 0.2$. With this method we can 
explain the observed ``factorisation breaking'' at the Tevatron, 
without introducing any new parameters. Our model predicts 
approximately 2--3\% diffractive events without phase--space cuts, 
e.g.\ in diffractive Z--production in $\mathrm{p}\bar{\mathrm{p}}$ 
at $\sqrt{s} = 1.8$ TeV we obtain 2.64\%, where data from D0 
implies approximately 1.44\% \cite{Abazov:2003ti}. 
Restricting the phase space 
in the event generation, by applying the cuts used in the 
experiments, further reduces the fraction of diffractive 
events, bringing our model closer to data. But the fraction 
of diffractive events is not the complete story. 
The model should also be able to describe particle 
spectra, and it is thus important to compare the kinematical 
distributions to data.

\begin{figure}[htb]
\centering
\includegraphics[scale=0.5]{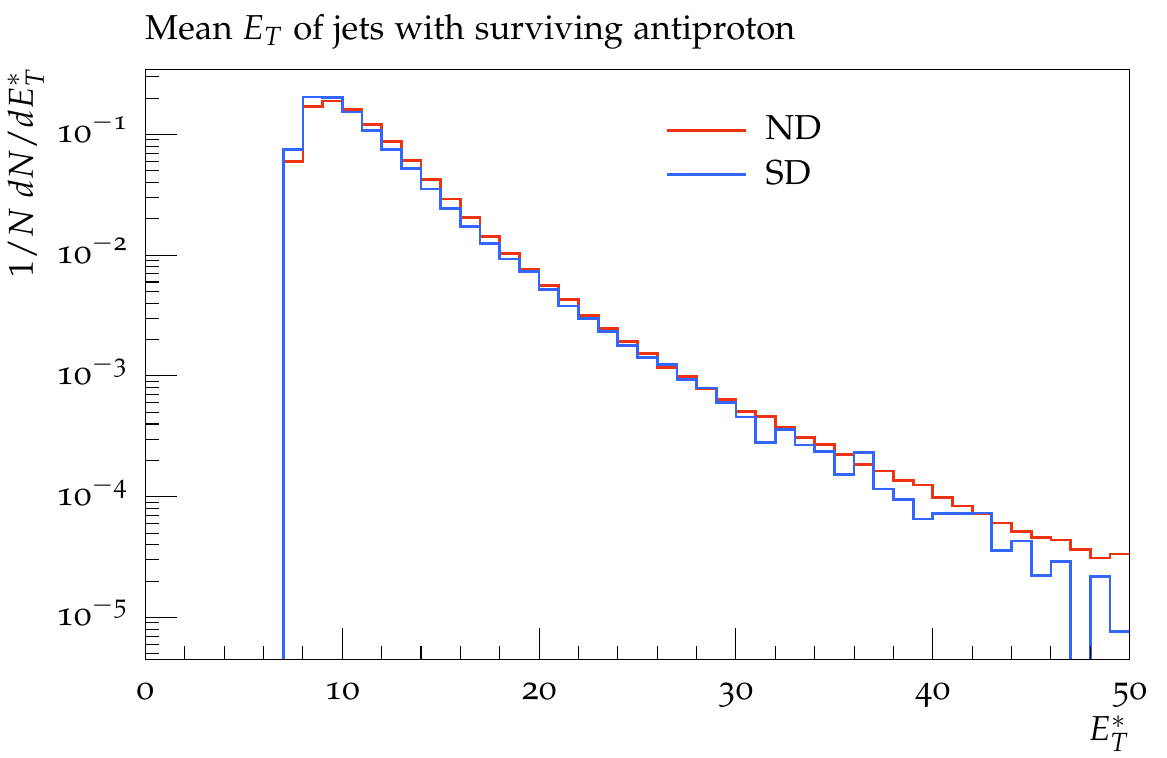}
\includegraphics[scale=0.5]{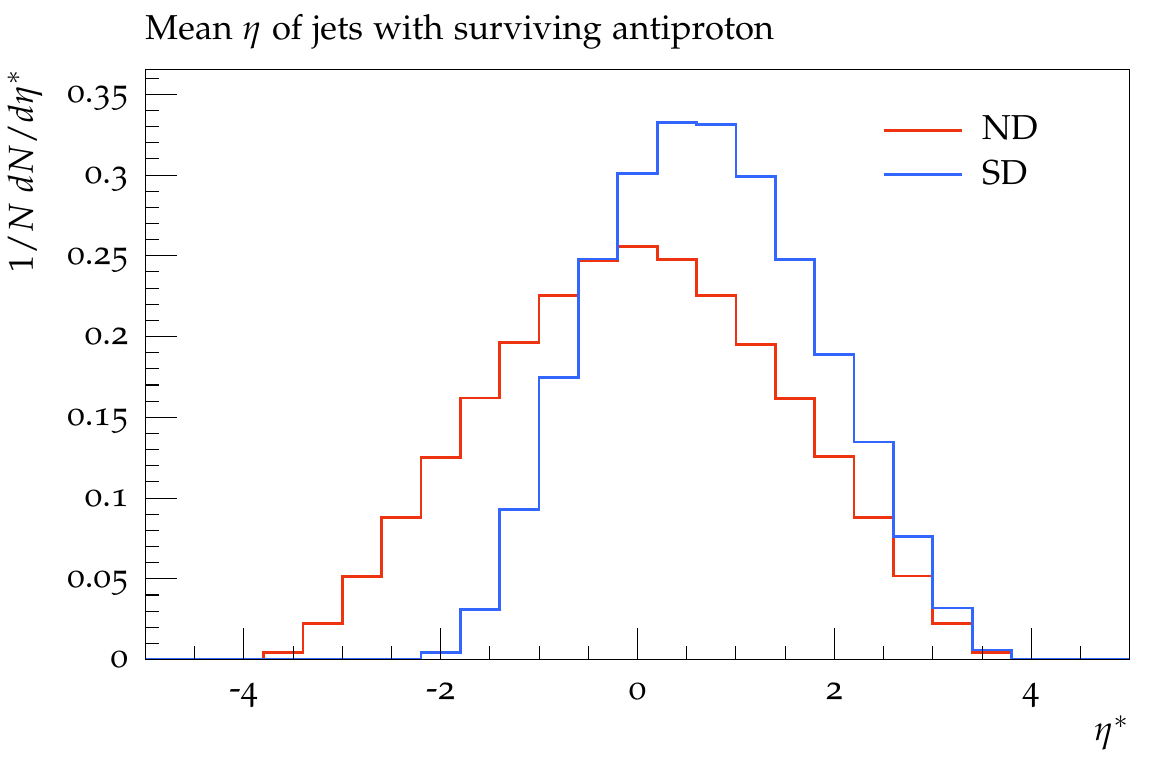}
\caption{\label{Fig:Kinematics}Kinematical distributions 
of diffractive dijet (SD) events compared to non--diffractive dijet 
(ND) events in $\mathrm{p}\bar{\mathrm{p}}$ collisions at 
$\sqrt{s} = 1.8$ TeV obtained with \textsc{Pythia}~8.}
\end{figure}
In Fig. \ref{Fig:Kinematics} we show some preliminary results 
obtained with the new model. We study diffractive dijet 
production at the Tevatron, 
$\mathrm{p}\bar{\mathrm{p}}\rightarrow \bar{\mathrm{p}} X, 
[X\rightarrow JJX']$ at $\sqrt{s}=1.8$ 
TeV. We show the mean $E_T$ and $\eta$ distributions, where the 
data obtained at the Tevatron showed significant differences 
compared to non-diffractive dijet events. SD data revealed  
a faster falloff in the mean $E_T$ distribution compared to 
ND events, and the events were  
shifted towards positive $\eta$, the proton direction. These 
differences implied a steeper $x$ dependence in the SD 
events than in ND events. 
Unfortunately, our model does not capture all of these effects. 
The SD events generated with the new model are boosted 
towards positive $\eta$ which is fine, 
but the falloff in $E_T^*$ is not significantly steeper 
than the ND distribution. Our model simply allows for too many 
high--$p_T$ events. While it may not solve all problems, 
we intend to develop a new description of 
the Pomeron flux to improve this spectrum. This should also 
improve the soft diffraction model implemented in 
\textsc{Pythia}~8.

\section{Conclusion}
We have presented a review of the soft diffraction models 
implemented in the general--purpose event generator 
\textsc{Pythia}~8. This soft diffraction machinery allows 
for QCD interactions and gives an decent description of 
diffractive phenomena. Comparisons to data shows that there 
is room for improvement in the default settings and a new 
parametrisations of the Pomeron flux is called for. 
A new model for hard diffraction has also been 
presented, now for the first time allowing for non--QCD 
processes as well has very high--$p_T$ QCD processes in 
diffractive systems. The model is successful in describing 
the RGSP, and diffractive fractions obtained with the 
model agrees reasonably with data. Particle spectra obtained 
with the model has been compared to the data, unfortunately 
not capturing all aspects of the data. Hence the required 
improvements and updates needed in the soft diffraction 
regime is also needed in the model for hard diffraction.  

\section*{Acknowledgements}

Work supported by the MCnetITN FP7 Marie Curie Initial 
Training Network, contract PITN-GA-2012-315877.


\begin{thebibliography}{99}

\bibitem{Sjostrand:2014zea}
T.~Sj\"ostrand {\it et al.},
Comput.\ Phys.\ Commun.\  {\bf 191} (2015) 159
[arXiv:1410.3012 [hep-ph]].

\bibitem{Ingelman:1984ns}
G.~Ingelman and P.~E.~Schlein,
Phys.\ Lett.\ B {\bf 152} (1985) 256.

\bibitem{Affolder:2000vb}
T.~Affolder {\it et al.} [CDF Collaboration],
Phys.\ Rev.\ Lett.\  {\bf 84} (2000) 5043.

\bibitem{Sjostrand:2006za}
T.~Sj\"ostrand, S.~Mrenna and P.~Z.~Skands,
JHEP {\bf 0605} (2006) 026
[hep-ph/0603175].

\bibitem{Navin:2010kk}
S.~Navin,
arXiv:1005.3894 [hep-ph].
                      
\bibitem{Donnachie:1992ny}
A.~Donnachie and P.~V.~Landshoff,
Phys.\ Lett.\ B {\bf 296} (1992) 227
[hep-ph/9209205].

\bibitem{Schuler:1993wr}
G.~A.~Schuler and T.~Sj\"ostrand,
Phys.\ Rev.\ D {\bf 49} (1994) 2257.

\bibitem{PythiaWeb}
\url{http://home.thep.lu.se/Pythia/pythia82html/Welcome.html}

\bibitem{Andersson:1983ia}
B.~Andersson, G.~Gustafson, G.~Ingelman and T.~Sj\"ostrand,
Phys.\ Rept.\  {\bf 97} (1983) 31.

\bibitem{Rasmussen:2015}
C.~O.~Rasmussen and T.~Sj\"ostrand,
in preparation.

\bibitem{Abazov:2003ti}
V.~M.~Abazov {\it et al.} [D0 Collaboration],
Phys.\ Lett.\ B {\bf 574} (2003) 169
[hep-ex/0308032].

\end{thebibliography}
\end{document}